\documentclass [article] {aa}

\usepackage{txfonts}

\usepackage{graphicx, epsfig}

\def\ltsima{$\; \buildrel < \over \sim \;$}
\def\lsim{\lower.5ex\hbox{\ltsima}}
\def\gtsima{$\; \buildrel > \over \sim \;$}
\def\gsim{\lower.5ex\hbox{\gtsima}}

\begin {document}
  \title{Multi-Frequency VLBI Observations of \\ NRAO 530 }
   \author{S.-W. Feng\inst{1,2}\thanks {\email{winfeng@shao.ac.cn}}
       \and Z.-Q. Shen\inst{1,3}
        \and H.-B. Cai\inst{1,2}
        \and X. Chen\inst{1,2}
        \and R.-S. Lu\inst{1,2}
        \and L. Huang\inst{1,2}
  }
  \institute{Shanghai Astronomical Observatory, Chinese Academy of Sciences, Shanghai 200030, China
  \and Graduate School of Chinese Academy of Sciences, Beijing 100012, China
  \and Joint Institute for Galaxy and Cosmology, SHAO and USTC, China
  }

\offprints{S.-W. Feng, winfeng@shao.ac.cn}
\date{Received $<$date$>$ / Accepted $<$date$>$}

\abstract {We report on VLBA observations of a $\gamma$-ray bright
blazar NRAO 530 at multiple frequencies (5, 8, 15, 22, 39, 43, and
45 GHz) in 1997 and 1999. These multi-epoch multi-frequency
high-resolution VLBI images exhibit a consistent core-dominated
morphology with a bending jet to the north of the core. The
quasi-simultaneous data observed at five frequencies (5, 8, 15, 22
and 43 GHz) in February 1997 enable us to estimate the spectra of
compact VLBI components in this highly variable source. Flat spectra
are seen in central two components (A and B), and the most compact
component A with the flattest spectral index at the south end is
identified as the core. Based on the synchrotron cooling timescale
argument, it is suggested that the observed inverted spectrum of
component C is caused by the free-free absorption (FFA), though the
synchrotron self-absorption (SSA) model cannot not be definitely
ruled out. While the SSA probably exists in component B, it is
likely that the same FFA would produce the spectral turnover toward
component B since the fitted FFA coefficients in both B and C
components are almost the same. If so, the projected size of such an
absorbing medium is at least about 25 pc.

By adding our new measurements to previous data, we obtain apparent
velocities of two components (B and E) of 10.2 c and 14.5 c,
respectively. These are consistent with that the emergence of VLBI
component is associated with the flux density outburst, i.e.
components B and E are related to strong $\gamma$-ray flares in
1994.2-1994.6 and 1995.4-1995.5, respectively. We further
investigate the spectral variability by making use of the
single-dish measurements covering a complete outburst profile from
mid-1994 to mid-1998. It shows a continuous increasing in the
turnover frequency during the rising phase, and a gradual decreasing
after passing the peak of the flare. Finally, we discuss the
equipartition Doppler-factor ($\delta_{eq}$) based on analysis of
magnetic field and obtain $\delta_{eq}$s of 3.7, 7.2 and 0.8 for
components A, B and C, respectively, which are consistent with bf a
larger flux density in component B, the non-detection of proper
motion in component C and a bent jet.

\keywords{galaxies: quasars: individual: NRAO 530 -- galaxies:
jets -- radio continuum: galaxies} }

\titlerunning{Multi-Frequency VLBI Observations of NRAO 530}
\authorrunning{Feng et al.}
\maketitle

\section{Introduction}
  \label{sect:intro}
NRAO 530 (PKS 1730-130) is an 18.5 mag QSO (Welch and
Spinrad~\cite{welch73}) with a redshift z=0.902
(Junkkarinen~\cite{Junkkarinen84}). It is variable at radio
wavelengths (Medd et al.~\cite{Medd72}; Marscher and
Marshall~\cite{Marscher79}) and is an emitter of X-ray (Marscher
and Broderick~\cite{Marscher81}). On the basis of its optical
variability (Pollock et al.~\cite{Pollock79}) NRAO 530 is also
classified as one of optically violent variables (OVVs).

\begin{figure}
   \begin{center}
   \includegraphics[width=85mm]{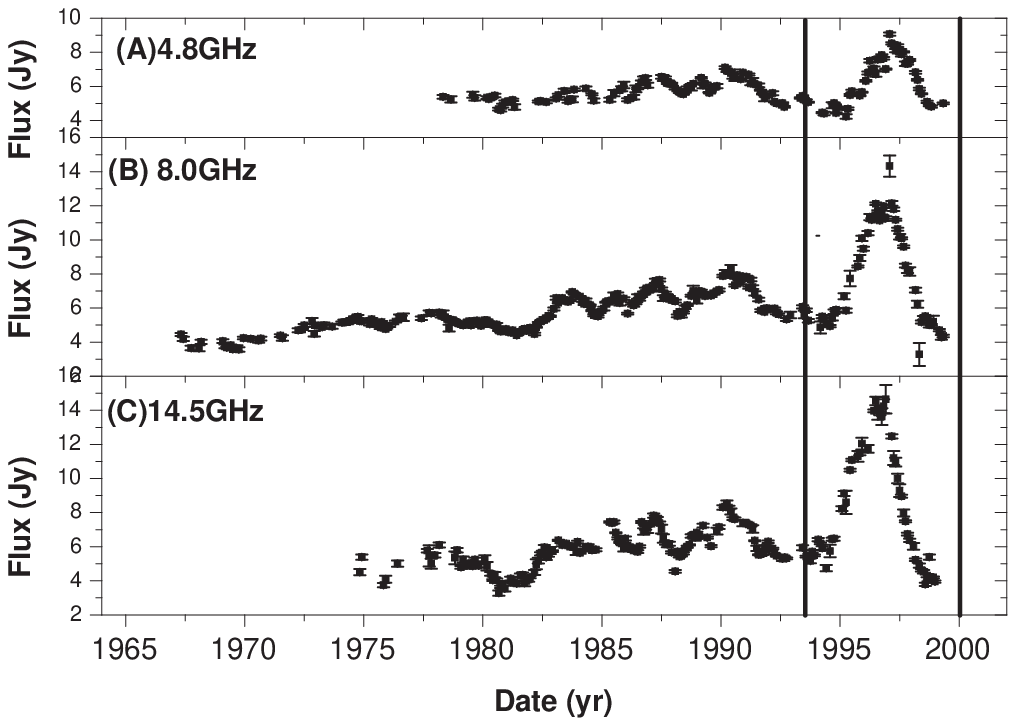}
   \includegraphics[width=85mm]{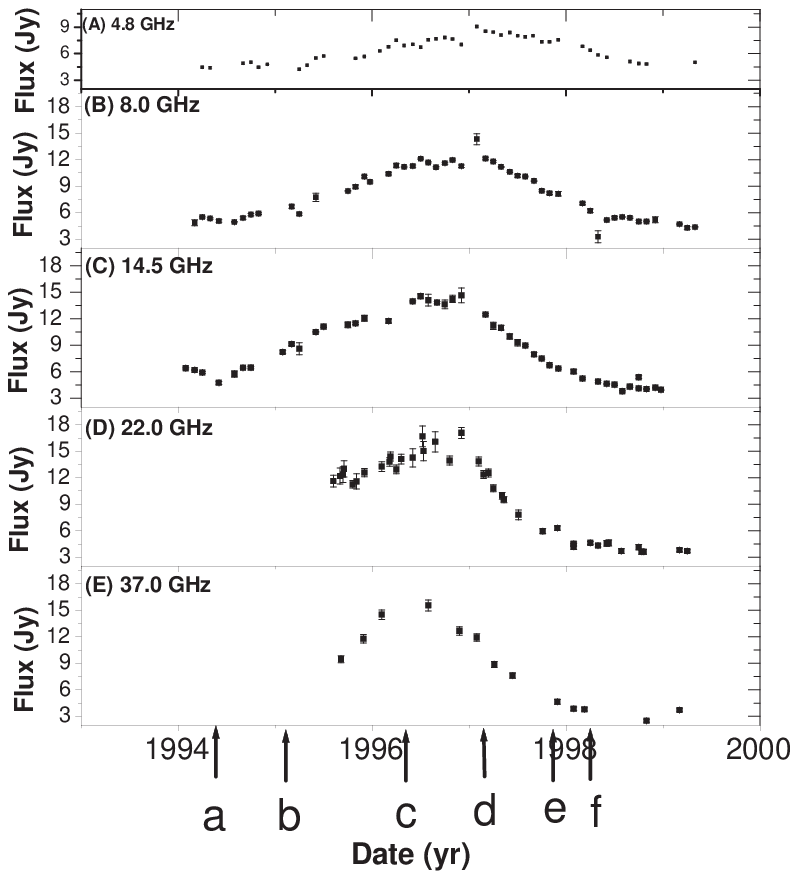}
\caption{ Top: light curves of NRAO 530 at (A)4.8 GHz, (B)8.0 GHz,
(C)14.5 GHz from UMRAO (University of Michigan Radio Astronomy
Observatory).  Bottom: light curves of NRAO 530 observed during the
period from 1994 to 1999 at (A)4.8 GHz, (B)8.0 GHz, (C)15.4 GHz from
UMRAO and (D)22 GHz, (E)37 GHz from 13.7-meter antenna at Metsahovi
Radio Observatory (Ter\"{a}sranta et al. ~\cite{Terasranta04}),
arrows with letters a, b, c, d, e and f indicate different epochs
plotted in Fig. 6.}
   \label{Fig:figure1}\end{center}
   \end{figure}

Figure~\ref{Fig:figure1} (Top) shows the flux density history at
4.8, 8.0 and 14.5 GHz monitored by the University of Michigan Radio
Astronomy Observatory (cf. Aller et al.~\cite{aller85}). It can be
seen that flux density changed greatly from 1994.5 to 1998.5. NRAO
530 underwent a dramatic millimeter flare beginning in 1994 with two
distinct flares nearly tripled the 90 GHz flux (Bower et
al.~\cite{Bower97}). EGRET reported increasing $\gamma$-ray flux
from the direction of NRAO 530 between 1991 and 1995 (Hartman et
al.~\cite{hartman99}). To investigate the relationship between these
flares (at radio and $\gamma$-ray) and the possible emergence of new
component, we present in this paper multi-epoch and multi-frequency
VLBI study of NRAO 530. We will discuss the kinematics of the jet
components, the possible absorption mechanism, and the spectral
variability.

Throughout this paper we adopt cosmological parameters $H_0$=71 km
s$^{-1}$ Mpc$^{-1}$, $\Omega_M$=0.27 and $\Omega_\Lambda$=0.73. This
results in a luminosity distance to NRAO 530 of $D_L$=5.8 Gpc, and
an angular-to-linear size conversion factor of 7.8 pc mas$^{-1}$.
The spectral index $\alpha$ is defined by
S$_{\nu}\propto\nu^{-\alpha}$.

%------------------------------------------------------------ Fig1: lightcurve
%-------------------------------------------------------------

\section{VLBI Observations and Data Reduction}

NRAO 530 was observed at 5, 8, 15, 22 and 43 GHz in February 1997
with the VLBA plus a VLA antenna. It was also observed 3 times in
1999 at 39, 43 and 45 GHz with the VLBA. Detailed parameters of
observations can be found in Table~\ref{Tab:table1}. The data were
recorded in VLBA format with a total bandwidth of 32 MHz (4 IFs) for
left circular polarization (LCP) (and also right circular
polarization (RCP) in 1997). The observed data were cross-correlated
with the VLBA correlator at Socorro, New Mexico of the USA.

%________________________________________ Table 1:
 \begin{table*}
 \centering
 \begin{minipage}{140mm}
  \caption[]{VLBI Observations of NRAO 530 }
%%Please Capitalize the First Letter of Each Notional Word in table's caption
  \label{Tab:table1}
%  \centering
  \begin{tabular}{ccccccc}
\hline\noalign{\smallskip}
\hline\noalign{\smallskip}
 $\nu$ &epoch &integration time &bandwidth &bit rate &polarization(s)&antennas    \\
 (GHz) &(year)&(min)            &(MHz)     &(bit)    &               &                 \\
  (1)  &(2)     &(3)       &(4)          &(5)     &  (6)         &  (7)          \\
\hline\noalign{\smallskip}
5.0  & 1997.10  & 20      & 32$\times$2 &1       & RCP, LCP    & VLBA+VLA1       \\
8.4  & 1997.10  & 20      & 32$\times$2 &1       & RCP, LCP    & VLBA+VLA1      \\
15.4 & 1997.12  & 20      & 32$\times$2 &1       & RCP, LCP    & VLBA+VLA1       \\
22.2 & 1997.12  & 20      & 32$\times$2 &1       & RCP, LCP    & VLBA+VLA1        \\
43.2 & 1997.12  & 20      & 32$\times$2 &2       & RCP, LCP    & VLBA+VLA1     \\
43.1 & 1999.32  & 17      & 32         &2        & LCP         & VLBA     \\
43.1 & 1999.40  & 17      & 32         &2        & LCP         & VLBA    \\
39.1 & 1999.42  & 19      & 32         &2        & LCP         & VLBA     \\
45.1 & 1999.42  & 18      & 32         &2        & LCP         & VLBA    \\
  \noalign{\smallskip}\hline
\end{tabular}
\\
Notes: (1) Observing frequency in GHz; (2) Observing epoch in yr;
(3) Total on-source time in minutes; (4) Recording bandwidth in
MHz; (5) sampling rate in bit; (6) Polarization mode, and (7)
Observing antennas.
\end{minipage}
\end{table*}
%________________________________________ Table 1

\begin{figure*}[htbp]
   \centering
    \includegraphics[width=145mm,angle=-90]{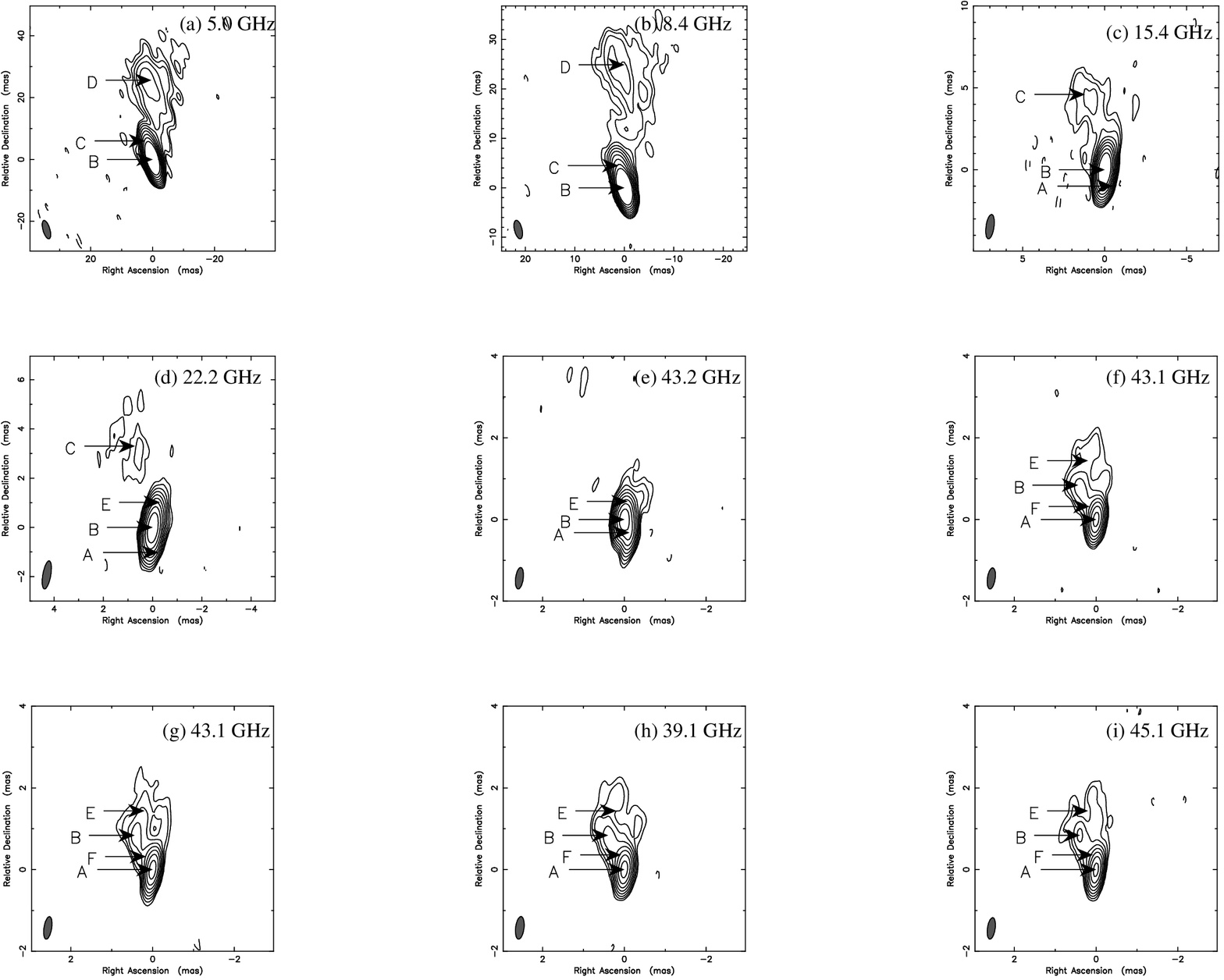}
\caption{The clean images of NRAO 530 (detailed parameters are
listed in Table~\ref{Tab:table2}).}
   \label{Fig:figure2}
 \end{figure*}

The post-correlation data reduction was carried out using the NRAO
AIPS and Caltech DIFMAP packages. We first applied amplitude
calibration (at frequencies $\geq$22 GHz atmospheric opacity
corrections were made) using the gain curves and system temperatures
measured at each station. Then fringe fitting was performed to
determine the instrumental delays first, and further to solve for
the residual delays, rates and phases. The data were then exported
into DIFMAP for self-calibration imaging. These visibilities were
time-averaged into 30-second bin with uncertainties estimated from
the scatter of data points within the 30 s interval. The data were
inspected for obviously bad points, most of which were near the
beginning of scan when telescopes were still slewing to the sources.
Images were made using the self-calibration and clean procedure for
several iterations. A 1-Jy point-source model was employed at the
start of each mapping process and the natural weighting of the data
was used for all nine images shown in Fig.~\ref{Fig:figure2}. The
detailed paraments of images are summarized in
Table~\ref{Tab:table2}. For a quantitative analysis of the structure
in these images, we did model fits to the calibrated visibility data
with elliptical Gaussian components. The results are listed in
Table~\ref{Tab:table3}. A maximum uncertainty of 15\% in the flux
density was estimated from the uncertainties of the amplitude
calibration and the formal errors of the model fits. The typical
uncertainties of the component positions and their sizes were
examined to be about 10\% of the fitted values. The brightness
temperature (column 11 in Table~\ref{Tab:table3}) of each component
was calculated using the following equation (i.e. Shen et
al.~\cite{Shen97}),

\begin{equation}
 T_{b}=1.22\times10^{12}S\nu^{-2}(\theta_{maj}\theta_{min})^{-1}(1+z)~ K,
\end{equation}
where $S$ is the flux density in Jy, $\nu$ the frequency in GHz, z
the redshift, $\theta_{maj}$ and $\theta_{min}$ sizes (FWHM) of
the major and minor axes in mas, respectively.
%------------------------------------------------------------ Fig2:images
%------------------------------------------------------------ Fig2
%________________________________________ Table 2
\begin{table*}[]
\centering
\begin{minipage}{140mm}
  \caption[]{Descriptions of VLBI images of NRAO 530 }
  \label{Tab:table2}
  \begin{tabular}{ccccccccc}
\hline\noalign{\smallskip}
\hline\noalign{\smallskip}
   & $\nu$ & Epoch      & S$_{peak}$      & Maj     & Min     & P.A.& Contours   \\
   &(GHz)  & (year)   & (Jy/beam) &(mas)   &(mas)   &(deg)&(mJy/beam)  \\
(1)&(2)    &(3)        &(4)        &(5)     &(6)     &(7)  &(8)        \\
  \hline\noalign{\smallskip}
(a)&5.0   & 1997.10    & 5.7      & 6.11    & 2.21 &16.1&2.9$\times$(-1,1,2,4,8,16,32,64,128,256)  \\
(b)&8.4   &1997.10     & 7.7      & 3.80    & 1.39 &15.5&2.8$\times$(-1,1,2,4,8,16,32,64,128,256)  \\
(c)& 15.4 &1997.12     & 8.0      & 1.49    & 0.49 &-8.0&8.4$\times$(-1,1,2,4,8,16,32,64,128,256)  \\
(d)&22.2  &1997.12     & 6.8      & 1.16    & 0.32 &-9.9&12.7$\times$(-1,1,2,4,8,16,32,64,128,256)  \\
(e)&43.2  &1997.12     & 4.5      & 0.53    & 0.19 &-7.3&10.0$\times$(-1,1,2,4,8,16,32,64,128,256)  \\
(f)&43.1  &1999.32     & 1.7      & 0.52    & 0.19 &-8.2&9.7$\times$(-1,1,2,4,8,16,32,64,128) \\
(g)&43.1  &1999.40     & 1.8      & 0.55    & 0.19 &-9.0&9.0$\times$(-1,1,2,4,8,16,32,64,128) \\
(h)&39.1  &1999.42     & 2.0      & 0.55    & 0.20 &-7.0&10.8$\times$(-1,1,2,4,8,16,32,64,128)  \\
(i)&45.1  &1999.42     & 1.7      & 0.53    & 0.19 &-8.3&10.2$\times$(-1,1,2,4,8,16,32,64,128)   \\
  \noalign{\smallskip}\hline
  \end{tabular}
\\
Notes: (1) Image index (see Fig. 2);
      (2) Observing frequency;
      (3) Observing epoch;
      (4) Peak flux density;
      (5), (6), (7) Parameters of the restoring Gaussian beam: the full
width at half maximum (FWHM) of the major and minor axes and the
position angle (P.A.) of the major axis.
      (8) Contour levels of the image.
      \end{minipage}
\end{table*}
%________________________________________ Table 2

%________________________________________ Table 3
\begin{table*}[]\centering
\begin{minipage}{140mm}

  \caption[]{NRAO 530 model parameters}
%%Please Capitalize the First Letter of Each Notional Word in table's caption
  \label{Tab:table3}
  \begin{center}\begin{tabular}{ccccccccccc}
\hline\noalign{\smallskip}
\hline\noalign{\smallskip}
   & $\nu$  & Epoch& Component&    S & r     &   $\phi$ & $\theta_{maj}$&$\theta_{min}$ &P.A.     & T$_{b}$ \\
   & (GHz)  &(year)&       & (Jy)  & (mas) &  (deg)    &(mas)          & (mas)         & (deg)   & (10$^{12}$K) \\
(1)& (2)  &  (3) & (4)      &(5)    &(6)    & (7)       & (8)           &(9)            & (10)           &(11)          \\
\hline\noalign{\smallskip}
(a)& 5.0&1997.10 &  B & 5.60  &   0  &   0        &0.78           &0.23           &   24.3   &   2.9     \\
   &    &        &  C & 0.60 &   4.02&   14.7    &3.13           &1.70            &    7.8   &   1.0E-2 \\
   &    &        &  D & 0.53  &  21.92&      0.4 &16.30          &6.12           &   13.2   &   4.9E-4 \\
(b)&8.4 &1997.10 &  B & 7.73  &     0 &        0  &0.35           &0.21           &    9.0   &    3.5 \\
  &     &        &  C & 0.64  &   3.22&      12.3 &4.07           &1.65           &   20.4   &   3.1E-3 \\
  &     &        & D & 0.33  &  22.89&       1.1 &13.60           &5.42           &   13.9   &   1.5E-4 \\
(c)&15.4&1997.12 & A & 0.91  &     0 &       0   &0.21           &0.21           &  ....    &   2.0E-1 \\
   &    &        & B & 8.01  &   0.38&      16.5 &0.22           &0.10           &   48.6   &   3.66 \\
   &    &        & C & 0.46  &   3.42&      10.2 &4.43           &1.70           &   22.4   &   5.9E-4 \\
(d)&22.2&1997.12 & A & 1.87  &     0 &        0  &0.26           &0.10           &   12.0   &   3.4E-1 \\
   &    &        & B & 6.73  &   0.29&      24.5 &0.21           &0.16           &   12.3   &   9.4E-1 \\
    &   &        & E & 0.12  &   0.73&       8.7 &0.14           &0.14           &  ...     &   2.9E-2 \\
    &   &        & C & 0.37  &   3.24&      14.5 &4.89           &1.68           &   13.1   &   2.1E-4 \\
(e)&43.2&1997.12 & A & 1.62  &     0 &        0  &0.13           &0.04           &    4.7   &   3.9E-1 \\
   &    &        & B & 5.76  &   0.35&      21.3 &0.17           &0.16           &  -42.3   &   2.6E-1 \\
   &    &        & E & 0.58  &   0.53&       2.4 &0.66           &0.28           &    4.3   &   3.9E-3 \\
(f)&43.1&1999.32 & A & 1.63  &     0 &         0 &0.08           &0.05           &   -0.3   &   5.1E-1  \\
   &    &        & F  & 0.34  &   0.18&      13.4 &0.19           &0.19           &  ...     &  1.2E-2 \\
   &    &        & B  & 0.28  &   0.84&      27.5 &0.59           &0.13           &   27.3   &  4.6E-3  \\
   &    &        & E  & 0.18  &   1.42&       2.8 &1.11           &0.47           &   17.7   &  4.3E-4 \\
(g)&43.1&1999.40 & A  & 1.86  &     0 &       0   &0.19           &0.04           &  -12.4   &  3.1E-1 \\
   &    &        & F  & 0.22  &   0.22&     20.8  &0.13           &0.13           &  ...     &  1.6E-2 \\
   &    &        & B  & 0.26  &   0.85&     27.9  &0.65           &0.18           &   18.3   &  2.8E-3 \\
   &    &        & E  & 0.22  &   1.33&      2.6  &1.22           &0.68           &   19.5   &  3.3E-4 \\
(h)&39.1&1999.42 & A  & 1.78  &     0 &       0   &0.07           &0.04           &   10.4   &  9.6E-1 \\
   &    &        & F   & 0.56  &   0.17&      12.2 &0.34           &0.09           &   45.5   & 2.8E-2 \\
   &    &        & B  & 0.31  &   0.92&      26.0 &0.63           &0.20           &   32.8   &  3.7E-3 \\
   &    &        & E  & 0.21  &   1.48&       2.8 &1.24           &0.54           &   23.6   &  4.8E-4 \\
(i)&45.1&1999.42 & A  & 1.65  &     0 &        0  &0.12           &0.02           &  -11.2   &  7.8E-1 \\
    &   &        & F  & 0.16  &   0.19&      18.5 &0.04           &0.04           &  ...     &   1.1E-1 \\
    &   &        & B  & 0.34  &   0.71&      27.4 &1.21           &0.12           &   18.3   &   2.7E-3 \\
    &   &        & E  & 0.16  &   1.29&      -1.6 &1.33           &0.21           &   11.0   &   6.5E-4 \\
 \noalign{\smallskip}\hline
  \end{tabular} \end{center}
Note: (1) Index of image (as shown in Fig.~\ref{Fig:figure2});
      (2) Observing frequency;
      (3) Observing epoch;
      (4) Component in each image;
(5), (6), (7), (8), (9), (10) Model parameters of Gaussian
component: S=flux density, r=angular separation and
$\phi$=position angle of component with respect to the origin
(0,0), full widths at half maximum (FWHM) of the major
($\theta_{maj}$) and minor ($\theta_{min}$) axes and the position
angle (P.A.) of the major axis;
      (11) Brightness temperature.
      \end{minipage}
\end{table*}
%________________________________________ Table 3

\section{Results}

\subsection{Components and Their Proper Motions}

\begin{figure*}
 \centering
  \includegraphics[width=160mm]{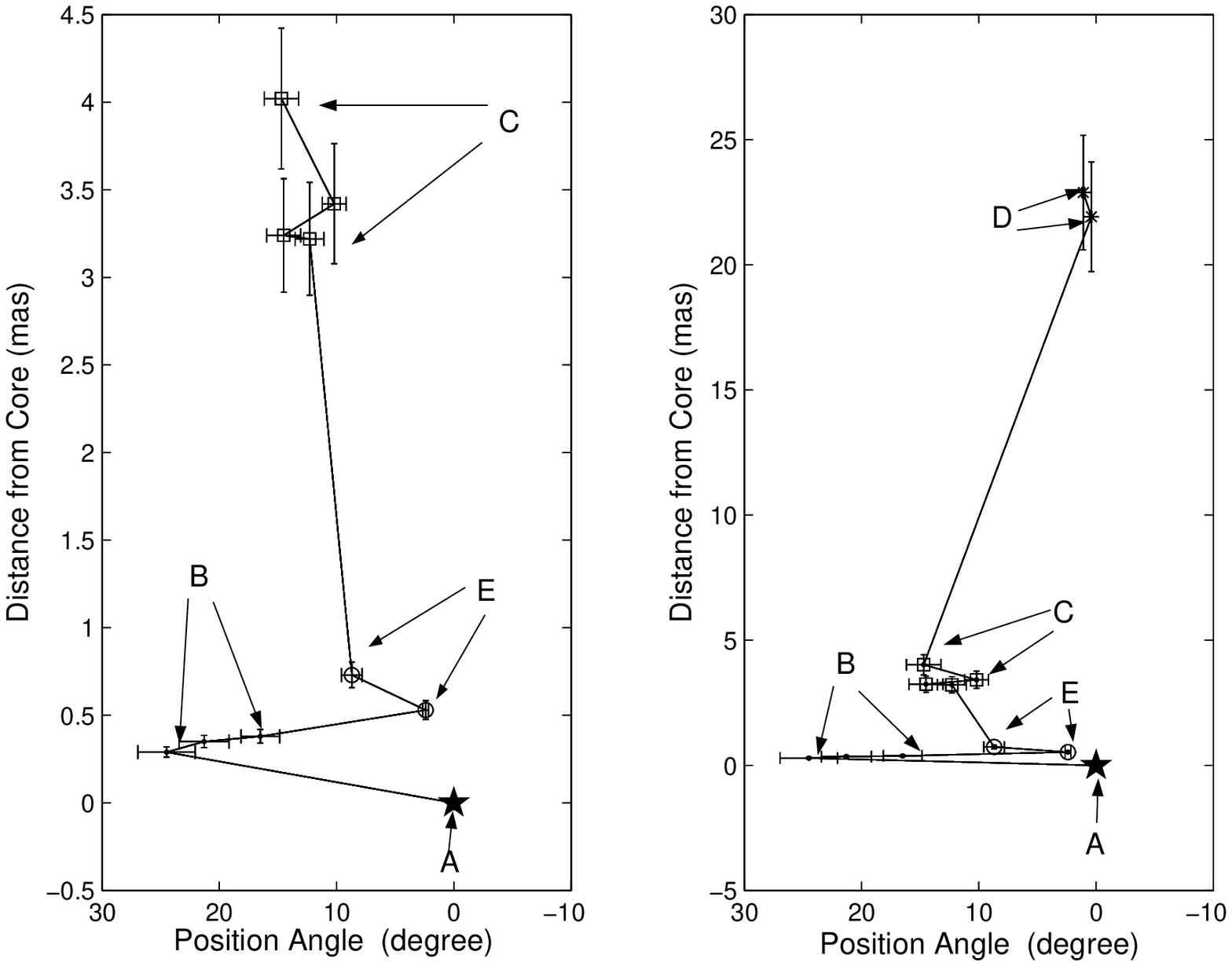}
\caption{The projected trajectory of jet components observed in
1997: (Left) within central 4 mas, (Right) within 20 mas, showing a
continuously bending jet moving away from the central core. A, B, E,
C and D represent different VLBI components and the solid lines are
connections of individual jet components.}
  \label{Fig:figure3}
\end{figure*}

The source shows different morphologies at different frequencies
with more details in the core region revealed at higher
frequencies (see Fig.~\ref{Fig:figure2}). The overall structure is
dominated by a compact core plus a jet like feature extending
approximately 5 mas from the core along a position angle (P.A.) of
15$^\circ$ which further bends toward the north.
Fig.~\ref{Fig:figure3} is the projected trajectory based on the
locations of the VLBI components in NRAO 530 observed in February
1997, inferring a continuously bending path of the jet emission.
Here, we assume the brightest and most compact component A (see
Table 3) as the core because it is located at the south end of
emission and has the flattest spectral index (see Sect. 3.2). The
same core identification is also adopted by Jorstad et al.
(\cite{Jorstad01}).

Component D at about 22-23 mas north of the core region was
detected at 5 and 8 GHz (Fig.~\ref{Fig:figure2} (a) and (b)). This
is probably associated with the knot emission, at about 25 mas
north of the core, seen from the 1.7 GHz VLBI observations (Bondi
et al.~\cite{Bondi96}). At about 3-4 mas from the core with a P.A.
of 10-15$^\circ$, component C was consistently detected at
frequencies of 5, 8, 15 and 22 GHz (Fig.~\ref{Fig:figure2} (a),
(b), (c) and (d)). This component was also seen by Tingay et al.
(\cite{Tingay98}), Shen et al. (\cite{Shen97}) and Jorstad et al.
(\cite{Jorstad01}).

The 1997 simultaneous observations at 43, 22 and 15 GHz showed a
consistent central core region (Fig.~\ref{Fig:figure2} (c), (d) and
(e)) which can be fitted with 3 components, A, B and E (Table 3).
The fitted relative separations from A component of 0.35 and 0.53
mas for components B and E, respectively at 43 GHz, are in a
reasonable agreement with the results from 22 and 15 GHz considering
their $>$1 mas angular resolutions (See Table 2). And it is also
possible that the $\sim$0.05$-$0.2 mas difference in the measured
core separation for components B and E is due to the opacity shift,
since the relative separation of VLBI component from the
self-absorbed core is frequency dependent. At 15 GHz, component E
seemed too weak to be detected (also refer to Fig.~\ref{Fig:figure5}
for its spectrum). These are comparable to the results obtained from
four 43 GHz images at epochs 1995.79 to 1996.90 (Jorstad et al.
\cite{Jorstad01}) and 86 GHz image at epoch 1995.38 (Bower et al.
\cite{Bower97}). Another four $\sim$43 GHz images
(Fig.~\ref{Fig:figure2} (f), (g), (h) and (i)) made in April-May
1999 can be well fitted with 4 components: components A, B, E and a
new component F at 0.18 mas with P.A. of 25$^\circ$.

\begin{figure}
 \centering
 \includegraphics[width=85mm]{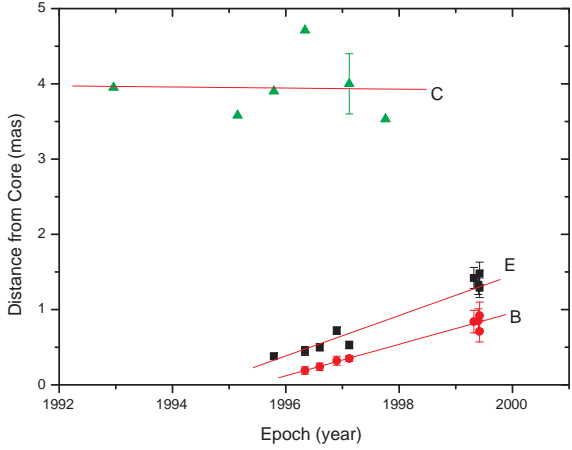}
\caption{Separation of components B, E and C from the assumed core
component A as the function of the observing epoch. }
 \label{Fig:figure4}
 \end{figure}

\begin{figure}

  \includegraphics[width=85mm]{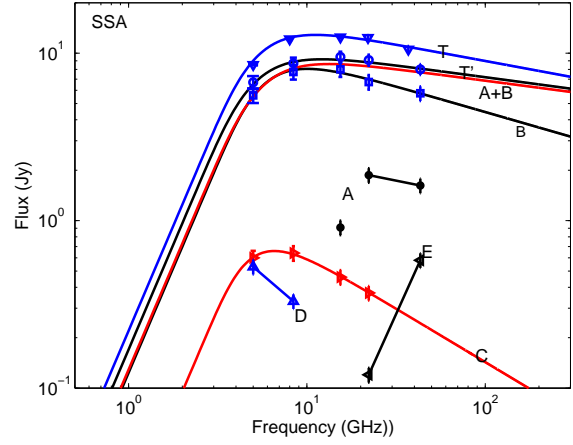}
  \includegraphics[width=85mm]{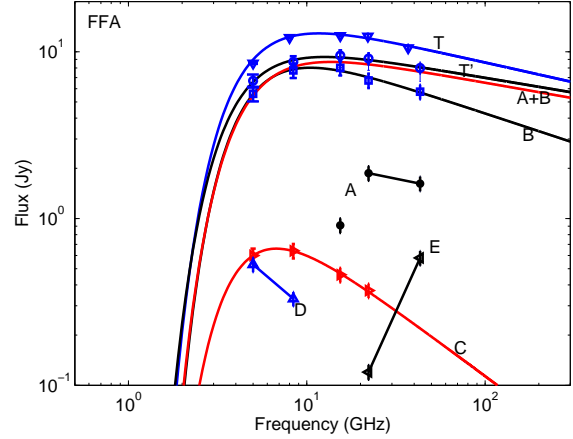}
\caption{Components spectral fit of NRAO 530 in February 1997 (top:
spectra are fit with the SSA, bottom: spectra are fit with the FFA}
  \label{Fig:figure5}
\end{figure}

By combining our new measurements of angular separations with those
published data in 1995 and 1996 (Jorstad et al. \cite{Jorstad01}) of
components B and E, we obtain proper motions of 0.21$\pm$0.02 and
0.30$\pm$0.02 mas yr$^{-1}$ (Fig.~\ref{Fig:figure4}), corresponding
to superluminal velocities of 10.2$\pm$0.8 c and 14.5$\pm$1.1 c for
components B and E, respectively. This infers the ejecting times for
components B and E to be around 1995.46 and 1994.76, respectively.
These two epochs just fall in the outburst phase shown in Fig. 1
(also see discussion in Sect. 3.3). EGRET reported the increasing
$\gamma$-ray flux densities from the direction of NRAO 530 during
two periods (1994.2-1994.6 and 1995.4-1995.5) with the average
fluxes of 64$\pm$26 pJy and 127$\pm$48 pJy (Hartman et
al.~\cite{hartman99}). As EGRET could not observe NRAO 530 with
dense time sampling and long integrations, the $\gamma$-ray
measurements have large uncertainties and can not rule out that
$\gamma$-ray flares were missed or that maximum of an ongoing flare
was really seen. According to the existing data (VLBA and EGRET) of
NRAO 530 we think that the emergence of components B and E are
closely following the $\gamma$-ray flares in NRAO 530. Activity of
$\gamma$-ray has been found to be correlated with infrared to X-ray
activity in 3C 279 (Maraschi et al.~\cite{Maraschi94}) and, to occur
before or during a radio and millimeter outburst in many sources
(Reich et al.~\cite{Reich93}).

For component C, we also plot its separation from the core
(Fig.~\ref{Fig:figure4}) as a function of time measured in 1992,
1995, 1996 and 1997 by Shen et al. (\cite{Shen97}), Jorstad et
al.~(\cite{Jorstad01}), Bower and Backer (\cite{Bower98}) and us
(this work), respectively. The best fitted proper motion is about
$-0.01\pm$0.13 mas yr$^{-1}$, corresponding to an apparent velocity
of $-0.35\pm$6.04 c. This is quite different from the apparent
velocity of 29$\pm$9$ h^{-1}$c (Hubble constant H$_{0}$=100 h km
s$^{-1}$ Mpc$^{-1}$, deceleration parameter q$_0$=0.1, and
cosmological constant $\Lambda=0$) reported by Jorstad et
al.~(\cite{Jorstad01}) (component D in their paper). Our new
estimate with more time coverage in data indicates that there is no
appreciable motion for component C. The previously reported large
apparent velocity of 29$\pm$9$ h^{-1}$c may be due to their limited
epochs of data and thus the large uncertainty. This discrepancy may
be related to the complex structure in component C (Fig. 2(c) and
(d)), which results in a large error in the determination of its
position. Such kind of diffused morphology could be due to internal
jet structure, blending effects of shocks or some sort of
interaction. For the reason to be discussed in Sect. 4, we think it
is mainly caused by the interaction of jet emission with the
surrounding medium.

%----------------------------------------------------- Figs 3 trajectory
%------------------------------------------------------------ Fig3

%------------------------------------------------------------ Fig4: proper motion
%------------------------------------------------------------ Fig4

%----------------------------------------------------- Figs 5 SSA+FFA
%________________________________________ Table 4
\begin{table*}\centering
\begin{minipage}{140mm}
\caption{Spectral fitting results with the SSA and FFA}
\label{Tab:table4}
\begin{center}
\begin{tabular}{cccccccc}
\hline\noalign{\smallskip} \hline\noalign{\smallskip}
Parameters  &\multicolumn{3}{c}{SSA}& \multicolumn{3}{c}{FFA}\\
 \cline{2-4}  \cline{6-8} \\
            &  B   &C    &A+B & & B   &C   &A+B   \\
\hline\noalign{\smallskip}
$\alpha$    &0.31  &0.64  &0.14&& 0.35 &0.79&0.19   \\
$S_{0}$(Jy) &0.12  &0.02  &0.13&& 21.77&4.38&15.59  \\
$\tau$      &148.31&159.90&101.40&& 23.33&20.88&21.89 \\
\noalign{\smallskip}\hline
\end{tabular}\end{center}
\end{minipage}
\end{table*}
%________________________________________ Table 4

\subsection{Components' Spectra}

For the quasi-simultaneous observations in February 1997, the
spectra of VLBI components (A, B, C and D) can be fit with
$S_\nu\propto\nu^{-\alpha}$ and the resultant optically thin
spectral indexes are $\alpha_A$=0.21, $\alpha_B$=0.31$\pm$0.07,
$\alpha_C$=0.58$\pm$0.01 and $\alpha_D$=0.91. Component A is
regarded as the core of NRAO 530 since it has the flattest spectral
index and it is at the south end of the overall emission. Component
E seems to show an inverted spectrum with a high turnover frequency
(greater than 43 GHz) in 1997. This is likely to be due to the
strong variability at 43 GHz as 1999 data had a three times lower
flux density. The fitted two-point spectral index is
$\alpha_E$=$-$1.5 between 22 and 43 GHz.

In order to discuss the possibly dominant mechanism for the
observed low frequency absorption in components B and C, both the
synchrotron self-absorption (SSA) and the free-free absorption
(FFA) are tried to fit the convex spectra with the following
formulae,
\begin{equation}
S_{\nu}=S_{0}\nu^{2.5}[1-{\rm exp}(-\tau_{s}\nu^{-\alpha-2.5})]
\end{equation}
for the SSA and,
\begin{equation}
S_{\nu}=S_{0}\nu^{-\alpha}{\rm exp}(-\tau_{f}\nu^{-2.1})
\end{equation}
for the FFA, where $\nu$ is the observing frequency in GHz, $S_0$
is the intrinsic flux density at 1 GHz in Jy, $\tau_f$ and
$\tau_s$ are the SSA and FFA coefficients at 1 GHz, respectively.
We also performed the fits to the combined component (A+B)
emission, which is believed to be represented by component B at
low frequencies (5 and 8 GHz) due to the limited resolutions (see
Table 3) and is resolved out at frequencies $\ge$15 GHz. All the
fitting results are listed in Table~\ref{Tab:table4} with the
fitting curves shown in Fig.~\ref{Fig:figure5}.

Both the SSA and FFA fit the spectra of components B and C equally
well. Thus, the spectral fit itself is not adequate to tell which
mechanism is mainly responsible for the absorption towards
components B and C. However, a short synchrotron cooling timescale
seems to suggest that FFA plays an important role in component C
(see Sect. 4). If so, that the FFA coefficients for components B and
C are almost the same (see Table 4) infers the presence of the FFA
by the same surrounding medium in component B. Then, one could
estimate that the absorbing material should cover at least about 25
pc in size to produce the same FFA towards components B and C which
were separated by about 3 mas in the sky in 1997. The interaction of
the surrounding interstellar medium causing FFA with the jet
emission in NRAO 530 can be used to explain the poorly determined
proper motion in component C (see Sect. 3.1).

 For comparison, we also plotted the spectrum of the sum of all the
VLBI components (labelled by T$^\prime$ in Fig.~\ref{Fig:figure5})
as well as the spectrum from single dish data (labelled by T in
Fig.~\ref{Fig:figure5}). The missing of flux density is persistently
seen in all five VLBI frequencies. This is likely due to the effect
of the resolution of the extended components. Therefore our
amplitude calibrations, including the atmospheric opacity
corrections, should be reasonable.
%----------------------------------------------------- Figs  6 variability
\begin{figure}
  \centering
   \includegraphics[width=85mm]{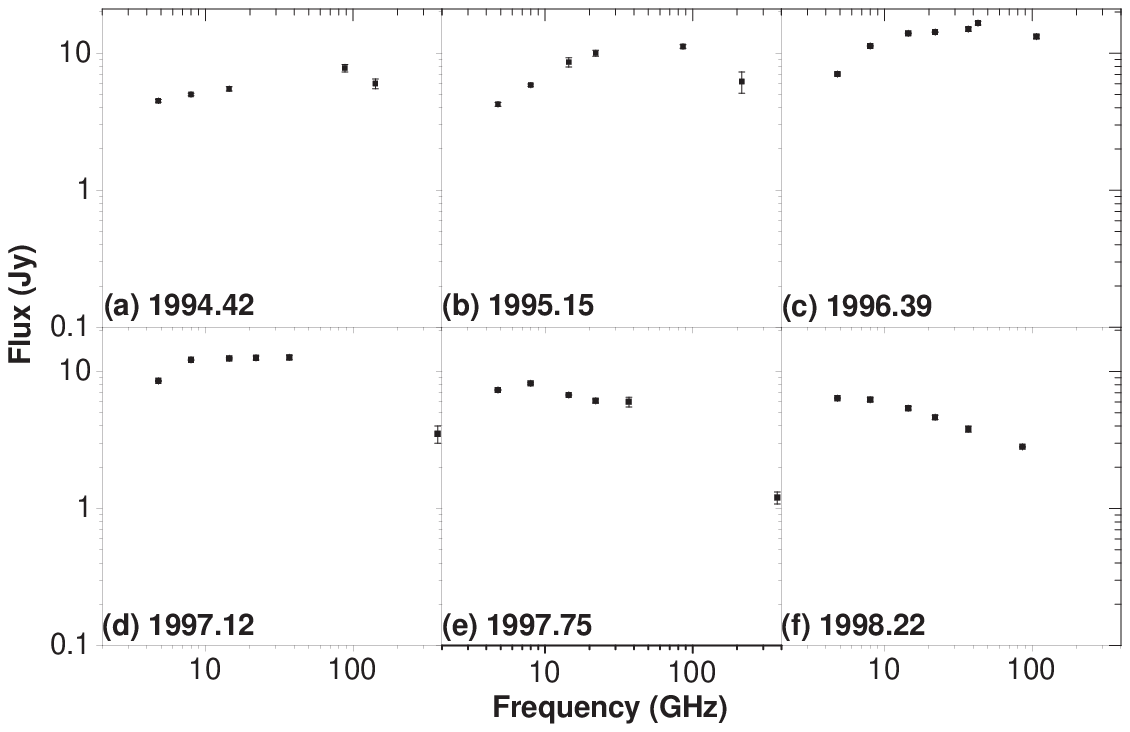}
\caption{Spectral variability seen in a flare from 1994.5 to 1998.5.
Flux densities at 43, 86, 88, 95, 107, 142, 150, 215 and 375 GHz are
listed in (Table~\ref{Tab:table5}). Letters a, b, c, d, e, f
represent different epochs indicated in Fig.~\ref{Fig:figure1}.}
  \label{Fig:figure6}
  \end{figure}
%------------------------------------------------------------ Fig6

\subsection{Spectral Variability}

A complete outburst profile can be traced from mid-1994 to
mid-1998 (between two vertical lines in Fig.~\ref{Fig:figure1}
(Top)). To analyze spectral variability during this radio outburst
we select six epochs: 1994.42 (a), 1995.18 (b), 1996.39 (c),
1997.12 (d), 1997.75 (e) and 1998.22 (f). Fig.~\ref{Fig:figure6}
shows the spectral distribution at these six epochs based on the
data points in Fig. 1 (at frequencies $\le$ 37.0 GHz) and in Table
5 (at frequencies $\ge$ 43 GHz). It is quite clear in
Fig.~\ref{Fig:figure6} that the turnover frequency is increasing
during the rising phase (represented by epochs (a), (b) and (c))
when the flux density is increasing (see Fig. 1).

From 1997.12 (d) the turnover frequency starts to decrease when
the total flux density decreases. This is because that high
frequency photons lose energy quickly with jet expanding and thus
the flux density at high frequency decrease too. In 1998.22 (f)
the spectrum returns to the level of quiescent phase.
%________________________________________ Table 5
\begin{table}
  \caption[]{Total Flux density data of NRAO 530 at $\nu\ge43$ GHz}
  \label{Tab:table5}
   \begin{center}
   \begin{tabular}{ccccl}
\hline\noalign{\smallskip} \hline\noalign{\smallskip}
$\nu$ &epoch     &flux&error &references                                           \\
(GHz) & (Year)         &(Jy)&(Jy)  &                                                 \\
\hline\noalign{\smallskip}
43    &1996.39&16.6 &0.5   &          Falcke et al.~\cite{Falcke98}          \\
86    &1995.18&11.2 &0.3   &          Krichbaum et al.~\cite{Krichbaum97}     \\
88    &1994.42&7.8  &0.5   &          Reuter et al.~\cite{Reuter97}           \\
95    &1996.82&12.6 &0.7   &          Falcke et al.~\cite{Falcke98}            \\
107   &1996.39&13.2 &0.4   &          Falcke et al.~\cite{Falcke98}              \\
142   &1994.42&6.0  &0.5   &          Reuter et al.~\cite{Reuter97}              \\
150   &1996.82&12.0 &1.2   &          Falcke et al.~\cite{Falcke98}             \\
215   &1995.18&6.2  &1.1   &          Krichbaum et al.~\cite{Krichbaum97}       \\
375   &1997.12&3.5  &0.5   &          Robson et al.~\cite{Robson01}              \\
375      &1997.26&2.7  &0.3   &          Robson et al.~\cite{Robson01}              \\
375      &1997.75&1.2  &0.1   &          Robson et al.~\cite{Robson01}              \\
\noalign{\smallskip}\hline
\end{tabular}\end{center}
\end{table}
%________________________________________ Table 5

\section{Discussion}

It is well known that the observed transverse velocity
($\beta_{app}=v_{app}/c$, $v_{app}$ the apparent velocity) is
related to the true velocity ($\beta$) and the angle to the line of
sight ($\theta$) by
$\beta_{app}=\beta\sin\theta/(1-\beta\cos\theta)$. Therefore, the
maximum angle ($\theta$) of 11.2$^\circ$ and 7.9$^\circ$ for
components B and E can be estimated from their corresponding
apparent velocities of 10.2 c and 14.5 c.

Doppler factor ($\delta$) is defined as
$\delta=\Gamma^{-1}(1-\beta\cos\theta)^{-1}$, where
$\Gamma=(1-\beta^2)^{-1/2}$ is the Lorentz factor. By means of the
above expression of $\beta_{app}$, we can derive
$\delta=\Gamma\pm(\Gamma^2-\beta_{app}^2-1)^{1/2}$. Then the Doppler
factor of component B is $\delta=\Gamma_{min}=10.2$ when the Lorentz
factor is $\Gamma_{min}=(1+\beta_{app}^2)^{1/2}$. In the same way we
compute the Doppler factor of another superluminal jet component E,
$\delta=14.5$.

For another scenario in which no intrinsic acceleration or
deceleration happened in the jet motion, i.e., jets are ejected with
a same velocity $\beta$ (or a constant Lorentz factor $\Gamma$), the
changes of apparent velocities would be mainly due to different
viewing angles (variable $\theta$). The observed fastest apparent
speed of 14.5c (for component E) in NRAO 530 will limit the constant
Lorentz factor to be at least of 14.5, and thus we use $\Gamma_{\rm
min}=15$.  Combining with the apparent velocity measurements, we can
obtain the viewing angles of $9.7^\circ$ and 1.6$^\circ$ for
component B with $\beta_{app}=10.2 \rm c$, and $4.9^\circ$ and
3.0$^\circ$ for component E with $\beta_{app}=14.5 \rm c$. The
corresponding Doppler factors are 4.0 and 26.0 for component B, and
11.3 and 18.7 for component E.

Estimates of the magnetic field in a radio source can be obtained in
two distinct ways. Firstly, by assuming the equipartition of energy
between the particles and the magnetic field, one can have (e.g.,
Pacholczyk \cite{pacholczyk70})
\begin{equation}
B^{\rm eq}=[4.5(1+k)f(\alpha,\nu_{1},\nu_{2})LR^{-3}]^{2/7}~,
\end{equation}
where B$^{\rm eq}$ in G is the magnetic field when the total energy
of magnetic field and particles has a minimum value (equipartition
between the energy of the particles and the energy of the magnetic
field), k is the energy ratio between the heavy particles and the
electrons (we used k=100), $f(\alpha,\nu_1,\nu_2$) is tabulated in
Appendix 2 of Pacholczyk (\cite{pacholczyk70}), $\alpha$ is the
optically thin spectral index, $\nu_1$ and $\nu_2$ are the lower and
upper cutoff frequencies of the synchrotron spectrum with typical
values of 10$^7$ and 10$^{11}$ Hz, $R=\frac{d_L\Theta}{2(1+z)}$ is
the radius of component in cm (here, $\Theta$ is the angular
diameter and d$_{L}$ is luminosity distance), L is the synchrotron
luminosity in erg~s$^{-1}$ and can be expressed as
L$=4\pi$$d_{L}^{2}\int_{\nu_1}^{\nu_2}S_{\nu}d\nu$ (here, S$_{\nu}$
is flux density at a frequency $\nu$). It should be mentioned that
B$^{\rm eq}$ is insensitive to the values of $k$ and the fraction of
the source's volume occupied by the magnetic field and the particles
which is auusmed to be 1. By assuming
$\int_{\nu_1}^{\nu_2}S_{\nu}d\nu \simeq S_m\nu_m$, we can re-write
\begin{equation}
B^{eq}\simeq4.0\times10^{-4}[f(\alpha,10^{7},10^{11})(1+z)^3d_{L}^{-1}\Theta^{-3}S_{m}\nu_{m}]^{2/7}~,
\end{equation}
where $\nu_{m}$ is the turnover frequency in GHz, $S_{m}$ is the
extrapolated optically thin flux density at the turnover frequency
in Jy,  and luminosity distance $d_L$ and angular diameter $\Theta$
are in Gpc and mas, respectively.

Secondly, assuming that the inverted spectrum is due to SSA, one can
estimate the magnetic field (i.e., Marscher \cite{Marscher83}),
\begin{equation}
 B^{syn}=10^{-5}b(\alpha)\Theta^{4}\nu^{5}_{m}S^{-2}_{m}\delta/(1+z)
\end{equation}
where $B^{syn}$ is the magnetic field in G, and b($\alpha$) is a
tabulated parameter dependent on the spectral index $\alpha$ (Table
1 in Marscher (\cite{Marscher83})). For component B, $\alpha$=0.31
gives b(0.31)$\approx$3.0.

From the SSA spectral fitting results for component B
($\Theta$=$\sqrt{\theta_{maj}\times\theta_{min}}$=0.2 mas, S$_m$=8
Jy, $\nu_m$=10 GHz) and $\delta=$1 (without consideration of
relativistically beaming effect), we obtained B$_{B}^{eq}\simeq580$
mG and $B_B^{syn}\simeq4\times10^{-2}$ mG. The time scale for an
electron to lose its energy through synchrotron radiation is
$t_{1/2}\simeq2.76\times 10^4~ ({\rm yr})~(\frac{B^{syn}}{\rm
1~mG})^{-1.5}(\frac{\nu_m}{\rm 1~GHz})^{-0.5}$. This gives a cooling
time scale of $1.1\times10^6$ yr for component B, while the
intrinsic cooling time scale in the source rest frame should be
greater when Doppler factor is larger than (1+z). In the same way,
we computed the magnetic field for component A,
$B_{A}^{eq}\simeq$560 mG and B$_A^{syn}\simeq8$ mG with S$_m$=1.87
Jy, $\nu_{m}$=22.2 GHz, $\Theta$=0.16 mas, $\alpha$=0.21 and
$\delta$=1 since it seems to have a (presumably synchrotron
self-absorbed) inverted spectrum at frequencies between 15 and 22
GHz (see Fig. 5).

The ratio of the particle energy density ($u_p$) to the energy
density of the magnetic field ($u_m$) can be expressed as (i.e.,
Readhead~\cite{readhead94})
$\frac{u_{p}}{u_{m}}=(\frac{B^{eq}}{B^{syn}})^{17/4}$. We get
$\frac{u_{p}}{u_{m}}\simeq6.9\times10^{7}$ and
$\frac{u_{p}}{u_{m}}\simeq5.5\times10^{17}$ for component A and B,
implying that both components A and B are particle dominated or that
their emissions are relativistically beamed. In the former case, for
component B this high ratio may account for the observed fast
decrease in the radio flux density (from 5.76 Jy to $\sim$0.3 Jy) in
two years even though its synchrotron cooling time could be much
longer ($\simeq1.1\times10^6$ yr). This is because that the high
ratio of particle to magnetic field energies may cause the inverse
Compton catastrophe (Readhead~\cite{readhead94}) and, inverse
Compton scattering of X-ray photons may further result in the
$\gamma$-ray emission (e.g., Hartman et al.~\cite{hartman92}).

In the latter case, as the equations for the magnetic field
(equations (5) and (6)) have different dependencies on the Doppler
boosting factor, thus we can obtain
\begin{equation}
\frac{B^{eq}}{B^{syn}}\simeq(\frac{B_{int}^{eq}}{B_{int}^{syn}})(\frac{\delta}{1+z})^{6+\frac{16\alpha}{7}}~~~,
\end{equation}
where $B^{eq}$ and $B^{syn}$ are the measured magnetic field,
$B_{int}^{eq}$ and $B_{int}^{syn}$ are the magnetic field estimated
in the source rest frame. When $B_{int}^{eq}\simeq B_{int}^{syn}$,
the Doppler factor in equation (7) is defined as the equipartition
Doppler-factor, $\delta^{eq}$. To derive equation (7), we have used
the following relations,
$\nu^{ob}_{m}=(\frac{\delta^{eq}}{1+z})\nu'_{m}$,
$S^{ob}_{m}=(\frac{\delta^{eq}}{1+z})^{3+\alpha}S'_{m}$ and
$\Theta^{ob}=(\frac{1+z}{\delta^{eq}})\Theta'$ (Singal and
Gopal-Krishna~\cite{singal85}), here $\Theta^{ob}$, $\nu^{ob}_{m}$
and $S^{ob}_{m}$ are angular diameter, inverted frequency and
corresponding flux density in the observer's frame, respectively,
and $\Theta'$, $\nu'_{m}$ and $S'_{m}$ in the source rest frame.
Then we got the equipartition Doppler-factors for components A and B
of $\delta^{eq}_{A}\simeq3.7$ and $\delta^{eq}_{B}\simeq7.2$,
respectively. These are comparable to the equipartition Doppler
factors of 7.0 at 10 GHz (Marscher et al. \cite{marscher77}) and 5.2
at 15 GHz (G\"{u}ijosa and Daly \cite{guijosa96}).

Similarly, for component C we estimate $B_{C}^{eq}\simeq36$ mG,
$B_{C}^{syn}\simeq29\times10^{3}$ mG,
$u_{p}/u_{m}\simeq7.4\times10^{-13}$, $\delta^{eq}_C\simeq0.8$ and
the synchrotron cooling time of $\sim$0.8 day, with the following
parameters: S$_m$=0.65 Jy, $\nu_m$=7 GHz, $\Theta$=2.5 mas,
$\alpha$=0.53 and $\delta$=1.0. As the magnetic field is strongly
dependent on measured component parameters, the high magnetic field
of component C may be due to its large size ($\sim 2.5$ mas).
Assuming that the equipartition is maintained in component C, we can
estimate a size scale of 0.64 mas for component C so that both the
calculated $B_{C}^{eq}$ and $B^{syn}_{C}$ are equal to 120 mG. The
resultant synchrotron cooling time would be 7.6 yr, still too short
for the SSA model. So, it is very likely that the FFA of the
surrounding medium dominates the observed absorption in component
though we cannot completely rule out the SSA effect.

As the Doppler-factors of jets are expected to increase with
decreasing viewing angle for the assumed constant Lorentz factor,
the equipartition Doppler-factors also should decrease
systematically along a bent jet given that the viewing angle is
increasing monotonically along the bending trajectory. This is
probably the reason that the $\delta_{eq}$ of component C is smaller
than the values of components A and B. However, the fact of
$\delta^{eq}_B>\delta^{eq}_A$ is consistent with that component B is
brighter than component A in 1997. For component B, the
Doppler-factor estimated from $\beta_{app}$ of
$\delta=\Gamma_{min}=10.2$ is comparable to the equipartition
Doppler-factor $\delta^{eq}_{B}\simeq7.2$, suggesting that most
likely jets have the minimum Lorentz factors in an equivalent state.
For component C, there is no appreciable proper motion (Section
3.1), which suggests a Doppler-factor of $\sim$1.0
($\delta=\Gamma_{min}=\sqrt{1+\beta_{app}^2}$). This is consistent
with the estimated equipartition Doppler-factor of 0.8, inferring
that it approaches energy equivalent state between particle and
magnetic field in the source rest frame. The fact that
$\delta^{eq}_C\simeq0.8$ is smaller than 1+z=1.902 for component C
in NRAO~530 implies that it is Hobble flow that results in the
observed departure from the energy equipartition in component C.

\section{Summary}

We presented the high-resolution VLBI images of NRAO 530 made at 5,
8, 15, 22, 39, 43, 45 GHz in 1997 and 1999. Combining our new
measurements with the past data we estimated superluminal motions of
components B and E at apparent velocities of 10.2 c and 14.5 c,
respectively. These are consistent with that components B and E were
ejected in 1995.46 and 1994.76 following the $\gamma$-ray flares of
127 and 64 pJy, respectively. Flat spectra are seen in central two
components (A and B), and the most compact component A with the
flattest spectral index at the south end is identified as the core.
We derive the equipartition Doppler-factor based on the equations of
(5) and (6), and obtain $\delta^{eq}$s of 3.7, 7.2 and 0.8 for
components A, B and C, respectively. These are consistent with a
larger flux density in component B, the non-detection of proper
motion in component C and a bent jet. The synchrotron cooling
timescale argues that the FFA is responsible for the inverted
spectrum of component C though the SSA effect can not be completely
ruled out yet. While for component B both the SSA and FFA could play
a role, we propose that the FFA may be at work given that almost the
same FFA coefficients were fitted for both components B and C. If
so, the absorbing material would have a projected size of at least
25 pc.

\begin{acknowledgement}
This research has made use of data from the University of Michigan
Radio Astronomy Observatory which has been supported by the
University of Michigan and the National Science Foundation. We thank
the anonymous referee for helpful comments and suggestions. This
work is supported in part by the National Natural Science Foundation
of China under grant 10573029. Z.-Q. Shen acknowledges the support
by the One-Hundred-Talent Program of Chinese Academy of Sciences.

\end{acknowledgement}

\bibliographystyle{aa} % style aa.bst

 \label{lastpage}
 \end{document}